\def\z{\zeta}
\documentclass[epj,referee]{svjour}
\usepackage{graphicx}
\begin{document}
\title{Zipf's law for fractal voids and a new void-finder}
%\subtitle{Zipf's law for fractal voids}
\author{Jos\'e Gaite
}                     % Do not remove
\institute{Instituto de Matem{\'a}ticas y F{\'\i}sica Fundamental,
CSIC, Serrano 113bis, 28006 Madrid, Spain}
\date{Received: date / Revised version: date}
\abstract{
Voids are a prominent feature of fractal point distributions but there
is no precise definition of what is a void (except in one dimension).
Here we propose a definition of voids that uses methods of discrete
stochastic geometry, in particular, Delaunay and Voronoi
tessellations, and we construct a new algorithm to search for voids in
a point set.  We find and rank-order the voids of suitable examples of
fractal point sets in one and two dimensions to test whether Zipf's
power-law holds.  We conclude affirmatively and, furthermore, that the
rank-ordering of voids conveys similar information to the
number-radius function, as regards the scaling regime and the
transition to homogeneity.  So it is an alternative tool in the
analysis of fractal point distributions with crossover to homogeneity
and, in particular, of the distribution of galaxies.
\PACS{
{05.45.Df}{Fractals}  \and
{02.50.-r}{Probability theory, stochastic processes, and statistics}
\and
{98.65.Dx}{Galaxy groups, clusters, and superclusters; large
scale structure of the Universe}
     } % end of PACS codes
} %end of abstract
\maketitle
\section{Introduction}
\label{intro}
As is well known, self-similar fractal sets can be characterized by a
power-law {\em number-radius} function (variation of the number of
points in a cluster with its radius).  Accordingly, tests for
fractality are made by measuring this function and fitting it to a
power law.  On the other hand, Mandelbrot in his seminal work on
fractals \cite{Mandel} introduced the concept of fractal holes, under
the greek word {\em tremas}, and studied their distribution, showing
that the distribution of one-dimensional holes (gaps) follows a simple
power law as well; namely, the number of gaps of length $U$ greater
than $u$ is $N(U > u) \propto u^{-D}$, where $D$ is the fractal
dimension. He also considered the higher dimensional generalization of
this distribution for those fractals with mathematically well-defined
holes, such as fractal carpets, sponges, foams, etc.  These fractal
sets have {\em topological codimension} one, so they are formed by
curves in two dimensions or surfaces in three dimensions (etc.) which
enclose empty holes.

Apart from Mandelbrot's general ideas, fractal holes have not been a
special object of study except in Cosmology \cite{Gaite}.  The
presence of large voids in the distribution of galaxies is known to
cosmologists since the late 70's, although their importance has only
been recognized recently. The distribution and properties of these
voids are now a subject of systematic study
\cite{cosmo,ElAd,AM,Hoyle1}.
%Ein,Kauffmann,ElAd,AM,Hoyle1}. 
Mandelbrot considered {\em
tremas} in the distribution of galaxies but, according to the
observational situation at the time, favored models with small voids
and, actually, introduced the concept of {\em lacunarity} to account
for this feature. At any rate, a fractal model of the distribution of
galaxies must have zero topological dimension, so the concept of hole
is not mathematically well defined. In fact, Mandelbrot's description
of the distribution of galaxies as having low lacunarity was based on
visual impression. The procedure to define voids in current
cosmological studies is algorithmic: given a galaxy catalog, one
designs an algorithm to find large empty regions sequentially (a {\em
void-finder}). Different void-finders provide different lists of
voids.  Presumably, to have sound statistics of voids in the galaxy
distribution or, generally, in abstract fractal sets one needs an
adequate definition of void.

We proposed in Ref.\ \cite{Gaite} to apply to voids in the
distribution of galaxies rank-ordering techniques
\cite{Zipf,Sornette}. A power-law rank order is Zipf's law \cite{Zipf}
and is equivalent to the law $N(U > u) \propto u^{-D}$.  We defined
voids in a particularly simple manner, namely, as having constant
shape (circles, squares, etc.) and showed that Zipf's law is fulfilled
by fractal voids independently of the chosen shape.  However, this
prescription for voids is far from satisfactory: often voids of
similar size touch each other and visually it seems that they should
be merged into a unique void of {\em irregular} shape.  This suggests
to improve the definition of void by considering voids of arbitrary
shape, in such a way that fractal voids still fulfill the Zipf law.

Regarding self-similar fractals in nature, we must take into account that
scale invariance has a limited range, between
a lower and an upper cutoff. For scales above the upper cutoff the
number-radius function $N(r) \propto r^D$ crosses over from $D < d$ to
homogeneity $D = d$ ($d$ is the embedding space dimension).  In
cosmology very large scale homogeneity is the basis of the standard
Friedmann-Robertson-Walker model, but the transition to homogeneity is
controversial~\cite{Piet-Marti}.

So the purpose of this work is: (i) to provide a definition of voids
in fractal sets of points (zero topological dimension) with transition
to homogeneity; namely, to devise a general algorithm to find voids in
fractal sets of points in $d$ dimensions showing a scaling (Zipf's law) 
that matches the scaling of voids in fractals in one
dimension; (ii) then, to compare the Zipf law for the rank-ordering of
voids with the standard calculation of the fractal dimension from the
number-radius function.  (iii) finally, to indicate how to apply our
conclusions to the voids in the galaxy distribution.

\section{Scaling of voids}
\label{sec:1}
It is easy to reach conclusions on the scaling of voids in
deterministic Cantor-like fractals constructed by a recursive
procedure, especially in one dimension \cite{Gaite}.  The fractal {\em
generator} is characterized by three independent numbers: $r,N,m$.
The first number, $r<1$, is the scaling factor: a $d$-dimensional unit
cube is divided into $1/r^d$ scaled cubes of size $r^d$.  One keeps
$N$ of the resulting $1/r^d$ cubes to repeat the process and removes
the others, leaving $m$ tremas, with ranks from $R_1=1$ to $m$ and
average size $\Lambda_1=(1-r^d N)/m$.  Of course, $1 \leq m \leq 1/r^d
- N$; the maximum number of tremas $m = 1/r^d - N$ is reached when
they are elementary cubes with no common facet.  In the second
iteration, there will be $mN$ tremas with ranks from $R_2=m+1$ to
$m+mN$ and average size $\Lambda_2=r^d(1-r^d N)/m$, etc.  The fractal
dimension $D=-\log N/\log r$ is independent of $m$.  In the $k$-th
iteration ($k \gg 1$), $R_k \propto N^{k-1}$ and $\Lambda_k
\propto (r^d)^{k-1}$, so the function $\Lambda(R)$ is a power law with
average slope $-d/D$.

A fractal set is partly characterized by its fractal dimension.
Dimension is a central concept in fractal geometry but it admits
several definitions \cite{Mandel}. Most important are the topological
dimension and the Hausdorff-Besicovitch dimension. Indeed, a fractal
is conventionally defined as a set for which the former dimension is
strictly smaller than the latter (which is simply called fractal
dimension). The fractal dimension of a self-similar fractal can be
calculated from the number-radius relation. The topological dimension
of a set is a topological invariant that can be determined from the
properties of its covering by small open sets \cite{Edgar}.  A set of
topological dimension zero is what we intuitively call a set of
points, a set of topological dimension one is what we intuitively call
a curve (or set of curves), etc. Here we use the topological {\em
codimension}, which is the dimension of the embeddding Euclidean space
($d$) minus the topological dimension.

In the construction of Cantor-like fractals, let us assume that every
removed trema gives rise to a unique void of the limit set, which is
true if the generator does not include any trema at the boundary so
tremas removed in different iterations cannot merge. This condition is
very natural in one dimension (merging in one dimension is limited,
anyway).  In $d > 1$ it ensures that the fractal set has topological
codimension one, because it contains the boundaries of tremas which
have themselves codimension one.  Typical fractals generated in this
form are the Sierpinski carpets \cite{Mandel}.  Therefore, by
identifying removed tremas with voids of the limit set, we have the
Zipf law $\Lambda(R) \propto R^{-d/D}$.  We remark that this law holds
on average, possibly with {\em log-periodic corrections}: within every
period the size of voids can be constant or change in an almost
arbitrary way.  Log-periodic corrections are the hallmark of discrete
scale invariance and naturally appear in Cantor like fractals
\cite{Sornette}.

If the fractal set does not have topological codimension one and, in
particular, if it has topological dimension zero (in $d > 1$), the
merging of generator tremas at different levels leads to the
complementary of the fractal set being connected. This demands a
criterium for separation of voids.

We must remark that the last number defining the generator, namely,
the number of tremas $m$, has no effect on dimension but characterizes
the {\em morphology} of the fractal. For given $r$ and $N$, and
therefore given fractal dimension, the largest $m = 1/r^d - N$
corresponds to the smallest tremas that the generator can have, and
viceversa. Therefore, $m$ is a measure of the size of tremas, that is,
of lacunarity.  We recall that the notion of lacunarity (from
the latin word ``lacuna'', meaning lake) was defined by Mandelbrot
\cite{Mandel}
as an intuitive measure of the size of tremas: fractals with larger
tremas (and the same dimension) are more ``lacunar''.  It is easy to
prove that the multiplicative factor in the Zipf law $\Lambda(R)
\propto R^{-d/D}$ is related with $m$ (in addition to $r$ and $N$)
\cite{Gaite}. In the log-log plot of the Zipf law, the multiplicative 
factor becomes an additive factor so it only affects the overall
position of the plot along the ordinate axis. This effect of
lacunarity is trivial.  However, there is no unique definition of
lacunarity. For example, a more elaborate definition of lacunarity
relates it with the three-point correlation function \cite{BB}. In
fact, higher order correlations contain morphological information that
goes beyond lacunarity, including information on the shape of voids.

\section{Discrete geometry methods and void-finder}
\label{sec:2}
To design a robust void finding algorithm, it is natural to rely on
concepts of discrete stochastic geometry (introduced in cosmology with
different objectives by Rien van de Weygaert and collaborators
\cite{Rien}).  The Delaunay tessellation of a $d$-dimensional point set
consists of a set of links between the points, forming simplices such
that their respective circumscribing $d$-spheres do not contain any
other point of the set. This tessellation is unique.  The Voronoi
tessellation is the dual tessellation formed by the centers of the
circumscribing spheres. Each Voronoi cell is the neighbourhood of a
point of the set, in the sense that the points of the interior of the
cell are closer to that point than to any other point of the set.

\begin{figure}
\centering{\includegraphics[width=7cm]{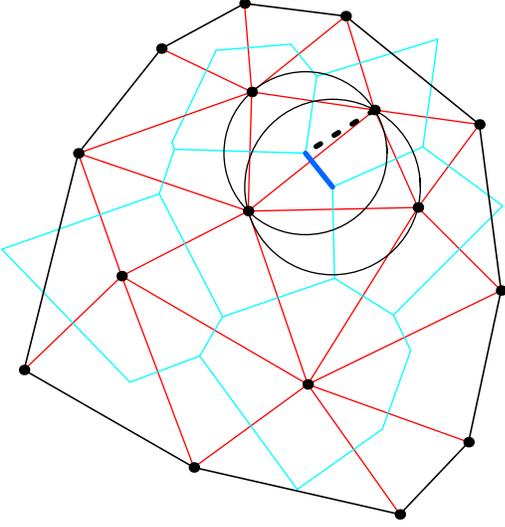}}
\caption{ 
Test set of 16 points showing the relevant discrete geometry
constructions: the Delaunay triangulation, the Voronoi cells of
interior points, and the circles circumscribed to two adjacent
triangles, with the segment linking their centers in boldface and the
radius of the smaller circle in dashed boldface. Our overlap parameter
$f$ is a bound to the ratio of the length of the segment linking their
centers to the length of the radius of the smaller circle.}
\label{f}
\end{figure}

The Delaunay and Voronoi tessellations are fundamental constructions
associated to a point set. In the search for voids, the defining
property of the Delaunay tessellation is obviously adequate: we can
consider each Delaunay simplex as an elementary void and define voids
by joining adjacent simplices according to some criterium.  The
natural criterium is given by the overlap of circumscribing spheres
corresponding to adjacent simplices being above a predetermined
threshold. This overlap criterium has been employed in various manners
in other void-finders that use spheres \cite{ElAd,AM,Hoyle1}. A simple
and efficient way to implement it is to demand that the distance
between the centers of two overlapping spheres be less than a given
fraction $f$ of the smaller radius (see Fig.\ \ref{f}). This criterium
generalizes the similar criterium in Ref.\ \cite{AM} by adding the
fraction $f$ as a free parameter (set to one in that reference).  The
advantage of having a free parameter is that one has some control on
the shape of the voids: for $f \ll 1$ we just have the elementary
voids, that is, the simplices of the Delaunay tessellation, but for $f
\simeq 1$ the voids adopt a ramified shape and the largest void may
percolate through the sample. When this happens, the percolating void
takes too large a fraction of the total size and imbalances the
statistics.  Actually, the imbalance effect takes place before
percolation and it is best to keep $f$ sufficiently small.

We have devised the algorithm for two-dimensional point sets but the
generalization to three (or more) dimensions is straightforward. To
begin the search for voids we need to estimate where we may find the
largest one, but we cannot measure the size of a void until it has
been found; so we look instead for the largest Delaunay triangle. The
algorithm consists of the following steps:

\begin{enumerate}
\item Construct the Delaunay triangulation and Voronoi tessellation
for the given point set.
\item Sort the triangles of the Delaunay triangulation by size and
select the largest one to begin to build the first void.
\item Grow the void by adding adjacent triangles (in this process, the
Voronoi tessellation is useful). A triangle is added if the overlap
criterium is met (for the chosen $f$). The set of all triangles found
constitutes the void.
\item Iterate by finding the largest triangle among the remaining ones
until they are exhausted.
\end{enumerate}

\section{Rank-ordering of voids and Zipf's law}
\label{sec:3}
Before applying our void-finder to fractals with transition to homogeneity,
let us consider the effect of this transition on the Zipf law. A transition to
homogeneity in a one-dimensional random Cantor-like fractal can be achieved by
joining by the ends several realizations of it.  For the sake of the argument,
let us assume that a single realization follows a perfect Zipf law $\Lambda_R
= \Lambda_1 R^{-\z}$. Then we have in 10 copies of the fractal, say, 10 voids
with size $\Lambda_1$ and ranks $R = 1, \ldots, 10$, 10 voids with size
$\Lambda_2$ and ranks $R = 11, \ldots, 20$, etc. So the sizes follow the law
$\log\Lambda_{10 n} = \log\Lambda_{1} + \z \log{10} - \z \log{(10 n)}$,
$\Lambda$ being constant between ranks $10 n - 9$ and $10 n$. This is a
stepcase with steps of exponentially decreasing width and linearly descending
ends with slope $\z$.  Relaxing the condition of an initial perfect Zipf law,
the steps become smooth, so we conclude that the effect of the transition to
homogeneity is the flattening of the Zipf law for small ranks and that the
width of the flattened portion measures the scale of homogeneity.

A coarse-graining procedure leads to a different picture of the
transition to homogeneity, which can be considered a bottom-up
picture, opposed to the preceding top-down picture. A fractal is
invariant under coarse-graining while in the scaling range, but
eventually the coarse-grained particles are no longer correlated and
correspond to a homogeneous distribution. Hence, we deduce that the
largest voids are the ones between the random (uncorrelated) coarse
particles corresponding to a homogeneous distribution. So it will be
useful to compare the small-rank voids with the ones of a random
distribution of particles.

\begin{figure}
%\centering
\includegraphics[width=8cm]{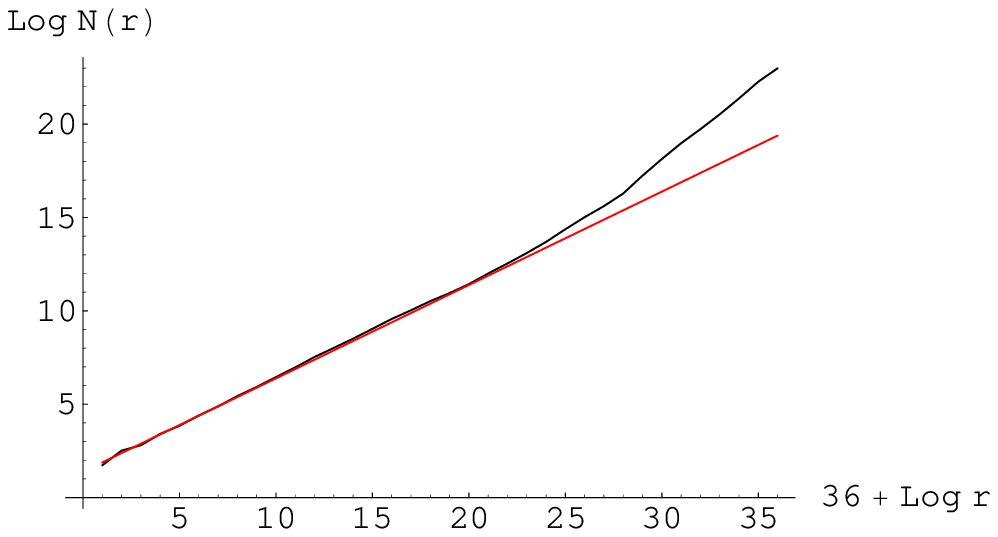}\hspace{2mm}
\includegraphics[width=8cm]{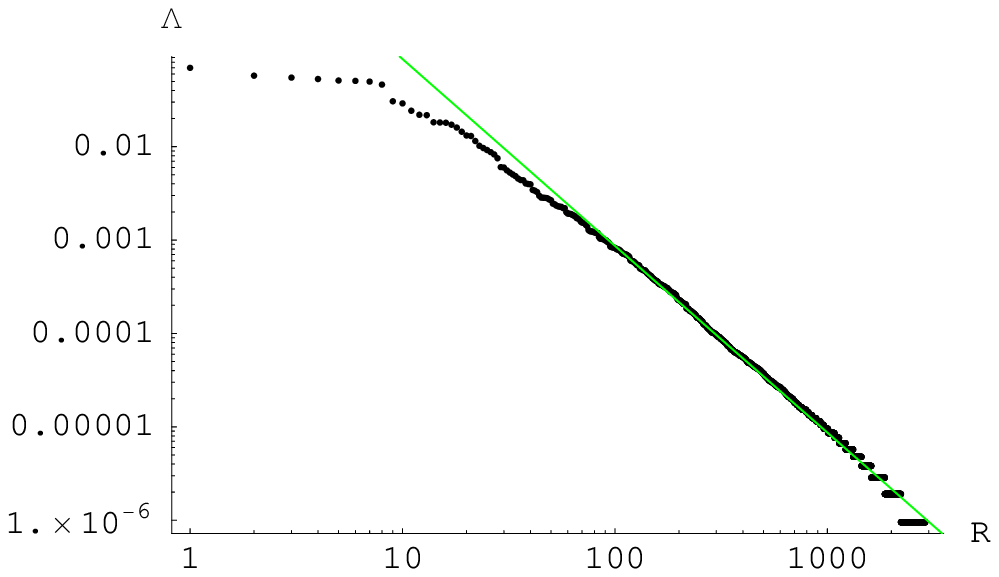}
%\vskip -.5cm
\caption{(Top) number-radius function for a $D=1/2$ fractal set in $d=1$
 (logarithms are to base $\sqrt{2}$ and the total size is normalized
 to unity), and (bottom) Zipf's law for the rank-ordering of its voids
 (gaps).}
\label{1d}
\end{figure}

To generate random fractals we use either random Cantor-like
algorithms or fractional Brownian motion (FBM) methods \cite{Voss}.
A fractional Brownian motion or its generalization to more than one 
dependent variable are not self-similar but {\em self-affine} 
\cite{Mandel,Voss}. However, appropriate sections are self-similar, 
so fractional Brownian functions constitute an adequate method to
generate random fractals, especially adequate when a precise control
of their fractal dimension is useful. We use the method of {\em
spectral synthesis}. We start from an array of wave-numbers, with a
given power-law power spectrum (with the appropriate Hurst exponent)
and random phases, and such that the amplitudes corresponding to
opposite wave-numbers are conjugate, to yield real data in real space.
Then we perform a fast Fourier transform to obtain these real-space
data. These self-affine data are then subjected to the appropriate
sections to construct the final self-similar fractal point set.

In Fig.\ \ref{1d} we have plotted the number-radius function and the
rank-ordering of voids for a FBM $D=1/2$ fractal with 2884 points in
$d=1$, with transition to homogeneity.  This transition appears in the
crossover of the number-radius function from $N(r) \propto r^{1/2}$ to
$N(r) \propto r$ on scales about fifteen times smaller than the total
size. This crossover has its counterpart in the flatness of the
rank-ordering of large voids (on similar scales).  Note that the
flattening is progressive, but the largest voids have approximately
the same size.

\begin{figure}
\centering{\includegraphics[width=7cm]{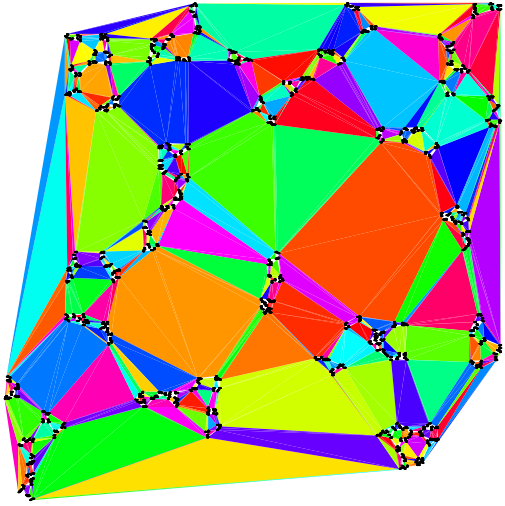}}
\caption{ Voids of a random Cantor fractal set with $D= 1$ and 12288
points (corresponding to overlap parameter $f=0.3$).}
\label{voids}
\end{figure}

\begin{figure}
%\centering
\includegraphics[width=8cm]{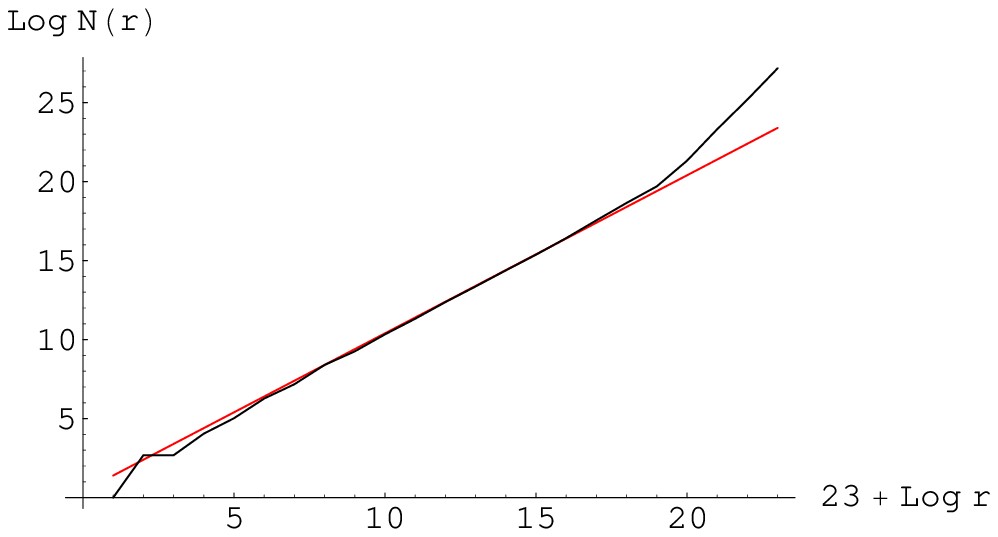}
\includegraphics[width=8cm]{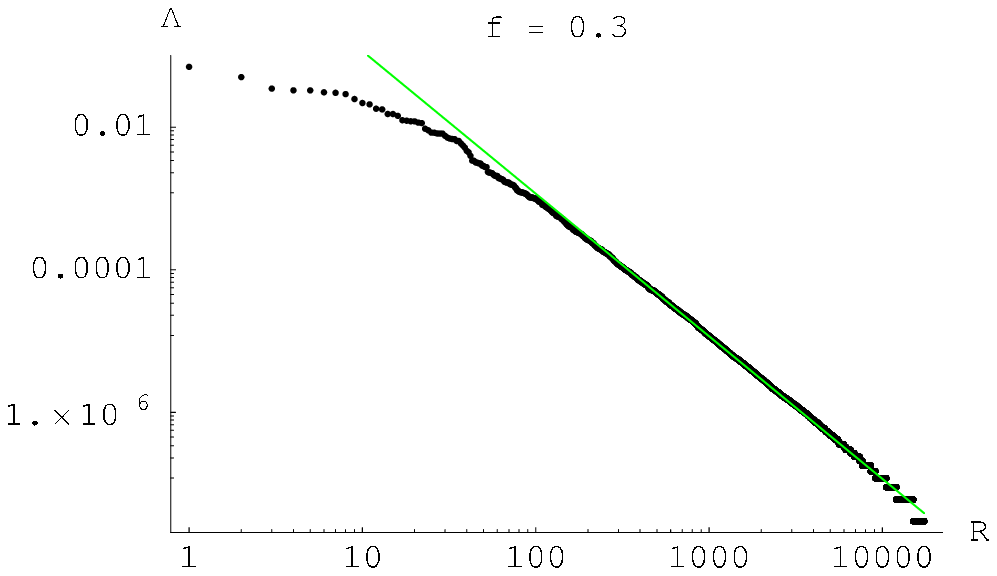}
%\vskip -.5cm
\caption{(Top) number-radius function of the $D=1$ fractal point set in
$d=2$ (Fig.\ \ref{voids}), and (bottom) Zipf's law for its voids
(corresponding to the displayed $f$).}
\label{2d}
\end{figure}

In two dimensions, we have run our void-finder on various fractal
point sets with different dimensions and various values of the overlap
parameter $f$.  As expected, Zipf's law is obtained; however, to get
the maximum scaling range (that is, similar to the scaling range of
the number function), the value of $f$ must be properly tuned.  To a
certain extent, this is due to the fact that one must keep away from
the percolation threshold.  Fig.~\ref{voids} shows the voids found in
a fractal with dimension $D= 1$ and 12288 points.  Fig.~\ref{2d} shows
its number-radius function and the Zipf law for voids.  We observe
that the transition to homogeneity has a similar representation in
both plots, namely, both graphs are approximately congruent.

We have determined the fractal dimension from either the values of
$N(r)$ or the sizes of voids.  From the number-radius function data we
estimate $D$ by a least-squares fit of the linear log-log relation.
The result depends on the portion we choose to fit, that is, on the
lower and upper cutoffs.  The best fit of $\log N(R)$ in Fig.\
\ref{2d} (top) is found for the range 10--15, yielding $D = 1.011 \pm
0.004$.  The ends of this range correspond to $r= 0.0055$ and $r=
0.031$, that is, a scale factor of only $(\sqrt{2})^{5} \simeq 5.7$,
0.but clearly the power law holds in a larger range: for example,
0.fitting the range 4--19 (scale factor 181) yields $D = 1.035 \pm
0.004$.  We apply a similar method to the rank-ordering of voids
0.(Fig.\ \ref{2d}, bottom), taking into account the discrete nature of
0.the rank.  The best fit is found for the range $512$--$8192$
0.($2^{9}$--$2^{13}$), yielding $2/D = 1.986 \pm 0.010$.  The ends of
0.this range correspond to $\Lambda^{1/2} = 0.0067$ and $\Lambda^{1/2} =
0.0.00042$, that is, a scale factor $16$, larger than the best-fit
0.scale factor for the number-radius function. However, the largest
0.scaling ranges are similar in both cases.

\begin{figure}
%\centering
\includegraphics[width=8cm]{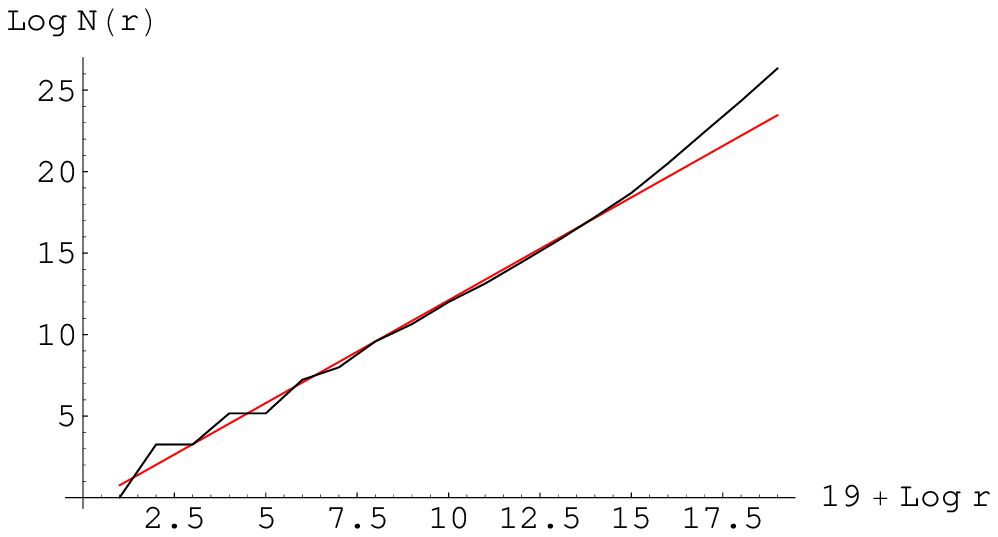}
\includegraphics[width=8cm]{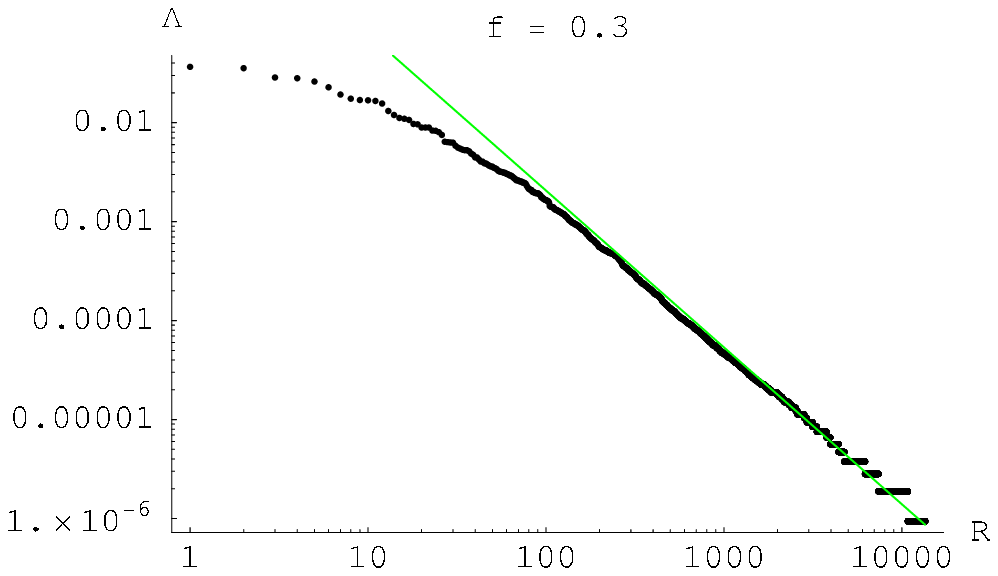}
%\vskip -.5cm
\caption{(Top) number-radius function of a $D=1.26$ fractal set in
$d=2$ with 9216 points, and (bottom) Zipf's law for its voids 
(corresponding to the displayed $f$).}
\label{2d2}
\end{figure}

For further illustration, we include another example in $d=2$, namely,
a fractal with dimension $D= 1.26$ and 9216 points.  Fig.~\ref{2d2}
shows its number-radius function and the Zipf law for voids.  The best
fit of $\log N(R)$ in Fig.\ \ref{2d2} (top) is found for the range
7--13, yielding $D = 1.27 \pm 0.02$.  We can also fit the slope of the
rank-ordering of voids (taking into account the discrete nature of the
rank) to give a quantitative measure of Zipf's law (Fig.\ \ref{2d2},
bottom).  For example, a fit of the range $256$--$8192$
($2^{8}$--$2^{13}$) yields $2/D = 1.51 \pm 0.02$.

\begin{figure}
\includegraphics[width=8cm]{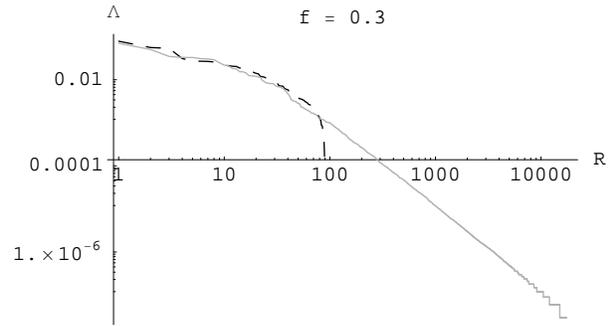}
\caption{Transition to homogeneity of the Zipf's law for the 
voids of the $D=1$ fractal point set in $d=2$ (continuous grey line)
compared with the voids of a random distribution of points having
small-rank voids of the same size (dashed line). The overlap fraction
used for both is $f=0.3$ (as displayed).}
\label{Zipf-Pois}
\end{figure}

As explained above, the precise way in which the scaling region of a
Zipf plot crosses over to homogeneity, namely, the way it flattens,
can be simulated by considering the corresponding random distribution
of coarse-grained particles. Indeed, looking at the pattern of voids
in Fig.~\ref{voids}, it is obvious that small voids between points in
a cluster are approximately independent of voids between clusters. The
largest coarse-grained particles correspond to clusters that are,
essentially, distributed randomly. So the largest voids must follow
the law of distribution of voids in a random point distribution. To
check it, we have applied our void-finder to a random distribution of
57 points in the unit square, which happens to have smallest-rank voids
of the same size as the smallest-rank voids of the fractal set in
Fig.~\ref{voids}. Then we have superposed the respective log-log plots
of rank-ordered voids in Fig.~\ref{Zipf-Pois} (joining the points
corresponding to individual voids).  Both plots agree sufficiently
well, within statistical errors, down to the rank where the scaling
regime begins. There the random point distribution has essentially no
more voids (except for a few small ones) and the corresponding line
falls abruptly.

\begin{figure}
\includegraphics[width=8cm]{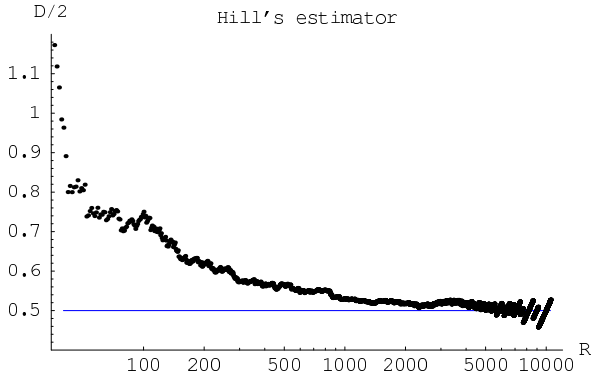}
\caption{Log-linear plot of the decrease with rank of the Hill
estimator for the Zipf law for the $D=1$ fractal voids, showing 
convergence towards $D/2 = 0.5$.}
\label{Hill}
\end{figure}

To derive the exponent of the Zipf law we can also apply
maximum-likelihood estimation through Hill's
estimator \cite{Sornette}
$$\frac{D}{2} = \left[\frac{1}{R} \sum_{i=1}^{R} \ln
\frac{\Lambda_i}{\Lambda_R}\right]^{-1},$$ taking data down to rank $R$.  The
result for our example is plotted in Fig.~\ref{Hill}.  The advantage
of this estimator is that it allows us to see how the small voids
progressively contribute to the measure of the value of $D$ and where
the fluctuations due to discreteness begin to make it less precise.

\section{Discussion}
\label{sec:4}
We must stress that the rank-ordering of voids and the corresponding
Zipf's law convey no more information than the number-radius function,
as regards scaling.  In fact, one may need to tune somewhat the
void-finder parameter to extract the same information. This is
probably due to the fact that the morphology of voids depends on the
type of fractal. The information on morphology (including features
like lacunarity) has value on its own but the existence of various
morphologies corresponding to the same scaling dimension poses
difficulties for void-finders. Especially, void-finders need to adapt
to the particular shape of voids of the given fractal.  Having tunable
parameters in the void-finder helps.  Clearly, these parameters must
only depend on relative magnitudes, like the one we have used ($f$),
but one has some liberty in their choice, nevertheless.

The rank-ordering of voids is likely to be useful for the analysis of
fractal distribution such that information on their voids is readily
available.  Regarding the distribution of galaxies, and considering
the great amount of information on its voids that is being compiled,
we infer that it is convenient to try to establish the Zipf law.
Moreover, we have demonstrated that the measure of the fractal
dimension $D$ provided by this law can reach similar accuracy to the
one given by the number-radius relation, which is the standard method
of measuring it. In addition, the rank-ordering of voids also provides
the scale of transition to homogeneity, and in a very intuitive
manner, since it is realted with the size of the largest voids.  This
information is already available in the catalogues of galaxy voids. In
contrast, the search for a scaling range in these catalogues fails
\cite{Gaite}. Hopefully, the use of simple and well-defined
void-finders, such as the one proposed here, and their use in the
compilation of more complete catalogues of galaxy voids will lead to
the observation of the scaling of voids and to independent measures of
the fractal dimension and the scale of transition to homogeneity of
the galaxy distribution.

\subsection*{Acknowledgments}
My work is supported by the ``Ram\'on y Cajal'' program
and by grant BFM2002-01014 of the Ministerio de Edu\-caci\'on y
Ciencia.

\end{document}